\documentclass[aps, prl, reprint, superscriptaddress]{revtex4-1}

\usepackage{graphicx}
\usepackage{bm}
\usepackage{amsmath}
\usepackage{hyperref}
\newcommand{\tise}{1\textit{T}-TiSe$_2$}
\newcommand{\stise}{1\textit{T}-TiSe$_2$\space}
\newcommand{\ev}{\varepsilon_\textrm{v}}
\newcommand{\ec}{\varepsilon_\textrm{c}}

\newcommand{\re}{\textrm{Re}}

\newcommand{\rmi}{\mathrm{i}}

\begin{document}
\title{Impact of Electron-Hole Correlations on the \stise Electronic Structure}

\author{G. Monney}\email{gael.monney@unifr.ch}\affiliation{D\'epartement de Physique and Fribourg center for Nanomaterials, Universit\'e de Fribourg, CH-1700 Fribourg, Switzerland}
\author{C. Monney}\affiliation{University of Z\"urich, Department of Physics, Winterthurerstrasse 190, CH-8057 Zu\"rich, Switzerland}
\author{B. Hildebrand}\affiliation{D\'epartement de Physique and Fribourg center for Nanomaterials, Universit\'e de Fribourg, CH-1700 Fribourg, Switzerland}
\author{P. Aebi}\affiliation{D\'epartement de Physique and Fribourg center for Nanomaterials, Universit\'e de Fribourg, CH-1700 Fribourg, Switzerland}
\author{H. Beck}\affiliation{D\'epartement de Physique and Fribourg center for Nanomaterials, Universit\'e de Fribourg, CH-1700 Fribourg, Switzerland}

\date{\today}

\begin{abstract}
Several experiments have been performed on \stise in order to identify whether the electronic structure is semimetallic or semiconducting without reaching a consensus. In this Letter, we theoretically study the impact of electron-hole and electron-phonon correlations on the bare semimetallic and semiconducting electronic structure. The resulting electron spectral functions provide a direct comparison of both cases and demonstrate that \stise is of predominant semiconducting character with some spectral weight crossing the Fermi level.
\end{abstract}

\maketitle
In noncorrelated materials, the semimetal or semiconductor character of a band-structure refers to the presence or absence of bands crossing the Fermi level. In correlated materials however, the electronic structure can be more subtle, showing coherent and incoherent parts with different spectral weights. \textit{Ab initio} band-structure methods reducing the Hamiltonian to a single-particle problem result in energy-momentum relations expressed by $\delta$-functions. These \textit{bare} bands are not directly accessible to measurements of correlated materials. Based on this consideration, the interpretation of the electronic structure should refer to the electron spectral function calculated via the self-energy. For \stise we have shown in a previous study \cite{Monney2012} that electron-hole fluctuations are strong even above the transition temperature, strong enough to influence the electronic structure around the Fermi level.

\stise is a layered quasi-two-dimensional (2D) compound with a commensurate 2 x 2 x 2 charge density wave (CDW) of critical temperature $T_\textrm{c} \approx 190$ K \cite{Salvo1976a}. The question of the nature of the gap (positive for a semiconductor or negative for a semimetal) above the transition temperature is widely debated. In 1976, based on electronic transport properties, Di Salvo \textit{et al}. \cite{Salvo1976a} claimed that \stise is a semimetal above the transition temperature. This assumption was confirmed in 1985 by Anderson \textit{et al}. \cite{Anderson1985} performing angle-resolved photoemission spectroscopy (ARPES), but in 2002, Kidd \textit{et al}. \cite{Kidd2002} found a very small indirect gap suggesting a semiconductor. In 2007, based on optical spectroscopy, Li \textit{et al}. \cite{Li2007a} disagreed, claiming that their measurements clearly reveal that the compound is metallic in the high-temperature normal phase. One year later, Rasch \textit{et al}. \cite{Rasch2008} challenged this conclusion with ARPES, defending that the analysis yields undoubtedly semiconducting behavior in \tise.

In this Letter, we evaluate the electron-hole susceptibility, calculate the acoustic phonon softening driving the transition to a lattice deformation in the CDW phase and calculate the influence of Coulomb and electron-phonon interactions on the electron spectral function for both the semimetal and the semiconductor bare-band structure. Starting from bare bands of either semiconducting or semimetallic structure we show that ARPES does not allow one to quantify the bare gap in the undeformed phase. However, there are differences between the two scenarios. First, contrary to the semimetal, where the fluctuations prepare a BCS-like transition, the semiconductor scenario shows typical features of a Bose-Einstein (BE) condensation. Second, the spectral function of the conduction-band has a coherent peak above the Fermi level in the semiconductor case but not in the semimetal case. Third, the calculation of the phonon softening as a function of temperature convincingly points towards a bare semiconducting electronic structure. These facts allow new insights to the semimetal versus semiconductor debate. The conduction-band coherent peak lays above the Fermi level, suggesting a semiconductor, but an incoherent part gives a non-negligible electronic contribution at the Fermi level explaining the semimetallic character claimed in some experiments.

Our electron-hole correlation model is based on work of Jerome \textit{et al.} \cite{Jerome1967} discussing the ground state of an excitonic insulator (EI) in the framework of Green functions. The EI phase is induced by electron-hole pairs coupled by Coulomb interaction that condensate at a critical temperature, forming quasiparticles called excitons. The initial gap of the system above the transition temperature to the EI phase determines the BCS or BE condensation type \cite{Bronold2006a, Phan2010, Kaneko2013}. The formalism we use here is similar to the one developed in previous papers \cite{Monney2012, Monney2012b}. Here we just mention the main steps of our calculations. The Hamiltonian contains a kinetic energy term composed by the dispersions of the topmost hole-like valence-band [$\ev(\bm{k})$] and the lowest electronlike conduction-band [$\ec(\bm{k})$] and an interaction term coupling the two bands via the Coulomb potential. The band dispersions are parabola fits to density function theory (DFT) dispersions \cite{DFTexpl} performed with the Wien2k package \cite{Schwarz2002}. In the hexagonal Brillouin zone (BZ) of \tise, the maximum of the valence-band is located at the center of the BZ, whereas the conduction-band has its minima at its border.

The extension of the EI model to fluctuations above the transition temperature introduces a two-particle interacting Green function $G_2$. It is computed from the zeroth order in the Coulomb potential $G_2^{(0)}$.  $G_2^{(0)}$ is a product of the Green functions for holes and electrons and $G_2$ is obtained via the Bethe-Salpeter (BS) equation
\begin{align*}
G_2&(\bm{Q},\bm{p},\bm{p'},\bm{w},z)=\delta_{\bm{p}\bm{p'}}G_2^{(0)}(\bm{Q},\bm{p},\bm{w},z)\nonumber\\
+&\rmi\frac{V_0}{\Omega}G_2^{(0)}(\bm{Q},\bm{p},\bm{w},z)\sum_{\bm{q}}G_2(\bm{Q},\bm{p}+\bm{q},\bm{p'},\bm{w},z),
\end{align*}
where $\bm{w}$ is the CDW wave vector between the centers of the valence and the conduction-band, and $\Omega$ is the volume of the unit cell. In order to solve the BS equation, we use a local potential which is constant in the reciprocal space ($V_0$). The coordinates $\lbrace\bm{Q}, \bm{p}, \bm{p'}\rbrace$ are the center-of-mass momentum and the relative momenta. Setting $\bm{Q}=\bm{0}$ is shifting the two parabola such that the centers match. Summed over the relative momenta $\bm{p}$ and $\bm{p'}$, the two-particle Green function $G_2$ has the properties of an electron-hole susceptibility [called $X(\bm{Q},z)$, where $z$ is the complex energy]. The approximation of the Coulomb potential by a constant $V_0$ in the reciprocal space allows one to write the susceptibility $X(\bm{Q},z)$ as a function of its noninteracting part $X^{(0)}(\bm{Q},z)$
\begin{align}
X(\bm{Q},z)=\frac{X^{(0)}(\bm{Q},z)}{1-\frac{V_0}{\Omega}X^{(0)}(\bm{Q},z)}.
\label{eq:X-X0}
\end{align}
The screening length corresponding to the magnitude of $V_0$ is comparable to the Thomas-Fermi screening length calculated for \stise via the plasma frequency \cite{Li2007a}. It is about one half of the nearest-neighbor distance, validating the local potential approximation, as discussed in a previous paper \cite{Monney2012}.
The noninteracting electron-hole susceptibility $X^{(0)}(\bm{Q},\omega) = \lim_{\delta\rightarrow 0}\left[X^{(0)}(\bm{Q},\omega-\rmi\delta)\right]$ is numerically computed with the Cuba package \cite{Hahn2005}. The self-energies of electrons, $\sigma_\textrm{v}$ and $\sigma_\textrm{c}$, and phonons $\sigma_\textrm{ph}$ are given by
\begin{align}
\sigma_\textrm{v}&(\bm{k},z_{\alpha})=\nonumber\\
&D^2\sum_{\bm{Q}}\int \frac{\text{d}\omega}{2\pi}\mathcal{X}(\bm{Q},\omega)\frac{N_{\textrm{B}}(\omega)+N_{\textrm{F}}(\varepsilon_\textrm{c}(\bm{k}+\bm{Q}+\bm{w}))}{z_{\alpha}+\omega-\varepsilon_\textrm{c}(\bm{k}+\bm{Q}+\bm{w})}\nonumber\\
\sigma_\textrm{c}&(\bm{k},z_{\alpha})=\nonumber\\
&D^2\sum_{\bm{Q}}\int \frac{\text{d}\omega}{2\pi}\mathcal{X}(\bm{Q},\omega)\frac{N_{\textrm{B}}(\omega)+1-N_{\textrm{F}}(\varepsilon_\textrm{v}(\bm{k}-\bm{Q}))}{z_{\alpha}-\omega-\varepsilon_\textrm{v}(\bm{k}-\bm{Q})}\nonumber\\
\sigma_\textrm{ph}&(\bm{Q},z_{\alpha})= g^2\left[X(\bm{Q},z_{\alpha})+X(\bm{Q},-z_{\alpha})\right].
\label{eq:selfs}
\end{align}
$N_{\textrm{F/B}}$ are the Fermi-Dirac and Bose-Einstein distribution functions, $D$ is the local Coulomb interaction $V_0/\Omega$ or the electron-phonon coupling $g$ (depending on whether we calculate the influence of the electron-hole or electron-phonon correlations on the electronic structure), $\mathcal{X}$ is the spectral function of the electron-hole susceptibility $X$ or of the acoustic phonon (not shown here, see Ref.\cite{Monney2012}) and $z_\alpha$ is the Matsubara frequency. The renormalized phonon frequency $\omega_r$ is given by the condition
\begin{align}
	\omega_r^2 - \omega_0^2 - \re\left[\sigma_\textrm{ph}(\bm{0},\omega_r+\rmi\delta)\right] = 0\textrm{, }\delta\to0,
	\label{eq:phonon}
\end{align}
where $\omega_0$ is the bare phonon frequency taken at the CDW wave vector at room temperature \cite{Holt2001} where the self-energy correction is small.
\begin{figure}
 	\includegraphics[width=\columnwidth]{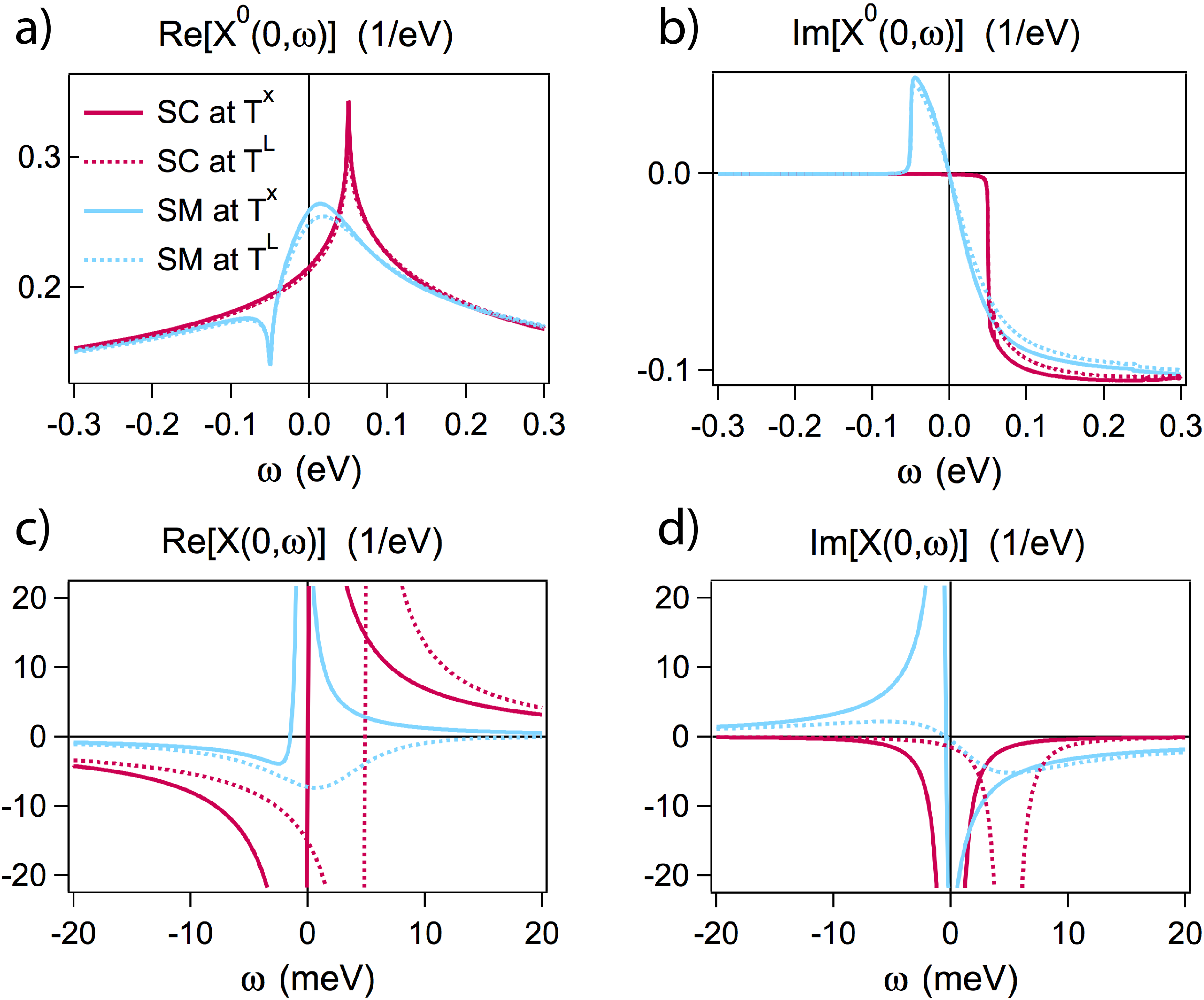}
 	\caption{\label{fig:1} Electron-hole susceptibilities as a function of energy for $\bm{Q}=\bm{0}$. In blue, the susceptibilities result from a semimetal with a negative gap of -50 meV. In red, the calculation is done for a semiconductor with a positive gap of 50 meV. Full lines are for $T^\textrm{X}$, the exciton condensation temperature. Dotted lines are for $T^\textrm{L}>T^\textrm{X}$, the temperature of the phonon softening (lattice instability). a) Real and b) imaginary part of the noninteracting susceptibility. c) Real and d) imaginary part of the interacting susceptibility.}
 \end{figure}
 
 \begin{figure*}
 	\includegraphics[width=\textwidth]{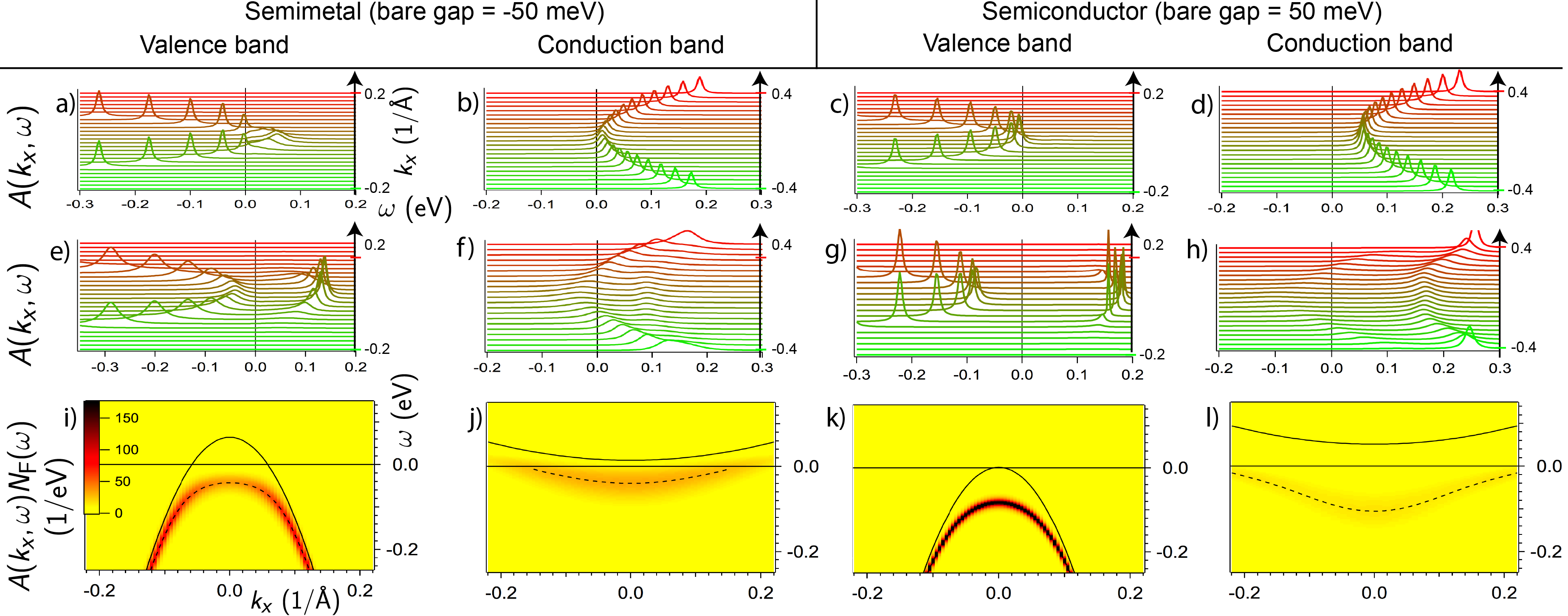}
 	\caption{\label{fig:2}(a-d) Electron spectral functions of the valence and conduction-bands for semimetal and semiconductor scenarios calculated via the self-energies due to electron-phonon correlations. (e-h) Same as (a-d), but with the contribution of both electron-hole and electron-phonon correlations to the self-energies. (i-l) Electron spectral functions corresponding to (e-h) cut by the Fermi-Dirac function.}
 \end{figure*}
Here we compare these quantities starting from a semiconducting or a semimetallic bare gap size. We set the chemical potential via the charge neutrality condition. The Coulomb potential $V_0$ in Eq. (\ref{eq:X-X0}) is chosen such that $X(\bm{0},0)$ diverges at the temperature $T^\textrm{X}$. Then, we calculate the phonon self-energy and the phonon renormalization (\ref{eq:phonon}) in order to find a temperature $T^\textrm{L}$ for which $\omega_r\to0$, i.e. the phonon softens, thereby inducing a lattice instability and the CDW phase. The parameters $V_0$ and $g$ are optimized such that the lattice instability temperature $T^\textrm{L}$ is equal to the experimental temperature of the CDW transition $T_\textrm{c}$. Finally, we calculate the electron spectral functions of the valence and the conduction-bands. The chemical potential is set again by the charge neutrality condition. For bare gaps between $-75$ meV to $75$ meV, the magnitude of $U=V_0/\Omega$ varies between 4.2 and 4.5 eV for semimetals and between 4.8 to 5.5 eV for semiconductors. The higher values for semiconductors are consistent with a less-screened potential. The electron-phonon coupling constant $g=0.5\pm0.1$ eV/\AA\; for both cases and compares well with Motizuki \cite{motizuki1986book} and Monney \textit{et al.} \cite{Monney2011a}. It turns out that $T^\textrm{L}>T^\textrm{X}$, as a consequence of Eq. (\ref{eq:phonon}) being satisfied for $\omega_r=0$ at a higher temperature than the divergence temperature of $X$ \cite{Monney2012}.

Figure \ref{fig:1} shows the electron-hole susceptibilities as a function of energy for the semimetal (blue curves, bare gap of $-50$ meV) and semiconductor (red curves, bare gap of $50$ meV) cases and for two different temperatures at $\bm{Q}=\bm{0}$. On the top, the real [Fig. \ref{fig:1}-a] and imaginary part [Fig. \ref{fig:1}-b] of the noninteracting susceptibility are presented, whereas on the bottom, the real [Fig. \ref{fig:1}-c] and the imaginary part [Fig. \ref{fig:1}-d] of the interacting susceptibility are shown.

The noninteracting susceptibility $X^{0}(\bm{Q},\omega)$ is a characteristic quantity calculated for CDW systems \cite{Johannes2006a, Rossnagel2011}. Its imaginary part [Fig. \ref{fig:1}-b] reflects the FS topology. The positive peak at negative energy (in blue) is characteristic of a semimetal band-structure and occurs at the energy of the negative gap. For the semiconductor band-structure (in red), $\textrm{Im}X^{0}(\bm{Q},\omega)$ is nonzero only for $\omega$ larger than the gap. Its real part is the relevant quantity for an instability. Here the peaks are too weak to induce a CDW instability (about two orders of magnitude weaker than peaks calculated for NbSe$_2$, for which the nesting scenario is still debated \cite{Johannes2006a,Soumyanarayanan2013}).

In the interacting susceptibility $X$ (Fig. \ref{fig:1}-c, d) a quasiparticle excitation occurs at very low energy. In a semimetal (in blue), the excitation peak grows with decreasing temperature until diverging at the critical temperature $T^\textrm{X}$ (solid line) forming a long-lived quasiparticle, the exciton (a condensed electron-hole pair bound by Coulomb interaction). If the lattice was rigid (no phonon softening), this divergence would induce the EI phase.  $X(\bm{0},z)$ has approximately the form of $1/(z^2-\omega'^2)$, where $\omega'$ is the excitation energy. As in the BCS model, the fluctuating excitations in the electron-hole continuum are stable only below the condensation temperature (forming Cooper pairs in superconductivity and excitons here). For the semiconductor (in red) however, the peak already diverges at a non-zero energy above $T^\textrm{X}$ (e.g., at $T^\textrm{L}$ the dashed curves). The exciton peak lays in the gap and $X(\bm{0},z)$ is well approximated by a bosonic Green function of the form $1/(z-\omega')$ preparing a BE condensation as the temperature decreases and the peak tends towards $\omega = 0$. But note that there is no exciton condensation here because the phonon softens at $T^\textrm{L}>T^\textrm{X}$.

Figure \ref{fig:2} presents the electron spectral functions in the semimetal (first column for the valence-band and second column the conduction-band) and the semiconductor (third column for the valence-band and fourth column for the conduction-band) cases for a gap amplitude of $50$ meV. They are calculated $10$ K above $T_\textrm{c}$.  In the first row, the valence-band (VB) and conduction-band (CB) spectral functions $A_\textrm{v/c}(k_x,\omega)$ are shown as energy dispersion curves (EDC), $x$ being the direction connecting the center of the BZ (location of the VB) to its border (location of the CB). They result from the valence- and conduction-band self-energies due to electron-phonon interaction ($D=g$ and $\mathcal{X}$ is the phonon spectral function). In the second row, the spectral functions are renormalized by the total self-energy i.e. the sum of the self-energy due to phonon interaction and due to electron-hole correlations where $D=V_0/\Omega$ and $\mathcal{X}$ is the spectral function of the electron-hole susceptibility $X$.  In the bottom row, the spectral functions displayed in false color are cut by the Fermi-Dirac function, showing only occupied states as accessible by ARPES. The solid lines show the bare bands fulfilling the charge neutrality condition and the dashed lines show the position of the Lorentzian fits to the EDC.

From the comparison between the first and the second row, we conclude that the dominant effect on the electronic structure is due to electron-hole correlations. For both semimetal and semiconductor scenarios, the valence-band spectral function is split into two parts around $k_x = 0$ [Figs. \ref{fig:2}(a) and \ref{fig:2}(c)]. A peak lays about 0.15 eV above the Fermi level. As a consequence, the initial holelike parabola has lost some spectral weight around its top. This feature is qualitatively similar for both scenarios. For the conduction-band however, there is a significant difference. For the semimetal [Fig. \ref{fig:2}(b)], the bare conduction-band is just shifted downwards and crosses the Fermi level but for the semiconductor [Fig. \ref{fig:2}(d)], there is a coherent peak above the Fermi level (about 0.15 eV) in addition to a weaker incoherent peak crossing the Fermi level.
If cut by the Fermi-Dirac function, the semiconductor coherent peak vanishes and the spectral functions look very similar in both scenarios. This similarity is a consequence of the valence-band spectral weight peak above the Fermi level [Figs. \ref{fig:2}(a) and (c)]. In order to conserve the number of occupied states (charge neutrality), the spectral weight lost in the valence-band has to be compensated by electrons of the conduction-band below the Fermi level. Thus, electron-hole correlations tend to transfer some spectral weight in the conduction-band below the Fermi level independently of the initial bare gap (semiconductor or semimetal). From these considerations, it becomes clear that the comparison with ARPES experiments does not allow one to decide about the nature of the initial bare gap.
\begin{figure}
 	\includegraphics[width=0.9\columnwidth]{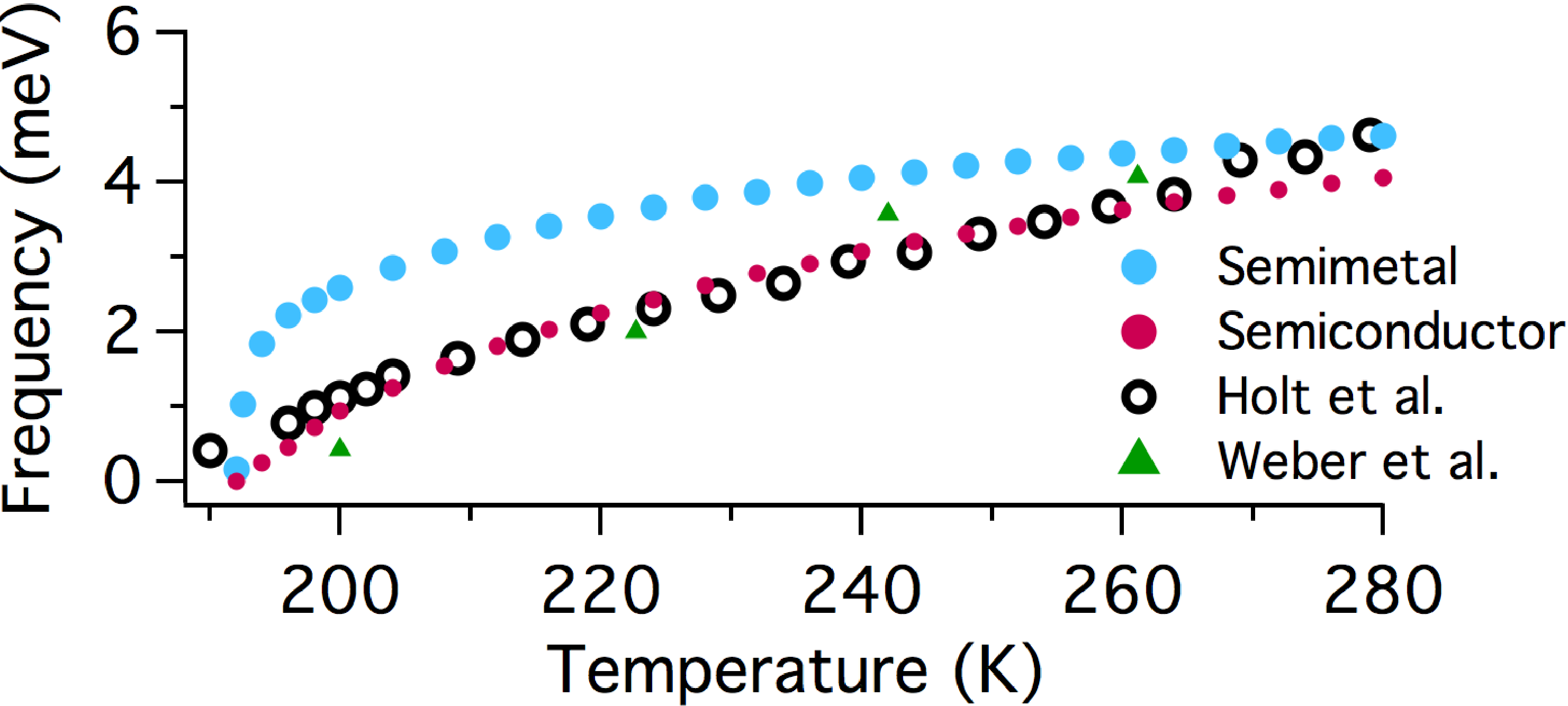}
 	\caption{\label{fig:3}Phonon softening calculated in the semimetal (blue dots) and semiconductor (red dots) scenarios. These calculations are compared to measurements of Holt \textit{et al.} \cite{Holt2001}(open black circles) and Weber \textit{et al.} \cite{Weber2011b}(green triangles).}
\end{figure}
The calculated softening of the phonon as a function of temperature is shown in Fig. \ref{fig:3}, together with the experimental curves. For a bare semimetal (blue dots), the renormalization is weak at high temperature and intensifies over a range of 10 K above the transition temperature. For a bare semiconductor (red dots), the renormalization is more regular. This behavior [given by Eq.\ref{eq:phonon}] is a direct consequence of the shape of the susceptibility $X$ in the semimetal or semiconductor scenario illustrated in Fig. \ref{fig:1}. The semiconductor scenario compares much better with experiments measuring the soft phonon mode in \stise by inelastic x-ray measurements \cite{Weber2011b} (open black dots) or x-ray thermal diffuse scattering \cite{Holt2001} (triangles). Note that both electron-hole and electron-phonon correlations cooperate in the phonon self-energy [Eq. \ref{eq:selfs}] to drive the transition. Though weaker, the effects of the electron-phonon correlations on the electron spectral functions tend to strengthen the effects of electron-hole correlations. This observation is agreement with the model of van Wezel \textit{et al.} \cite{VanWezel2010b} and confirmed experimentally by Porer \textit{et al.} \cite{Porer2014} claiming that both the electron-hole and electron-phonon coupling plays a significant role in \stise and that both effects cooperate to drive the CDW transition.

Our calculations provide a first direct comparison of the debated semimetal versus semiconductor scenarios above the CDW transition temperature in \tise. The results suggest a new way of considering the problem, more subtle than the conventional discussions in terms of bare $\delta$-functions bands. On the one hand, as suggested by the phonon renormalization which is directly proportional to the electron-hole susceptibility, this susceptibility should result from a semiconductor band-structure. The electron spectral function is also influenced by the electron-hole susceptibility. The renormalized conduction-band shows some spectral weight crossing the Fermi level. These facts are not contradictory because this spectral weight is an incoherent part while the coherent peak is positioned above the Fermi level.

This picture provides a reconciliation of the recent Letters \cite{Rasch2008,Li2007a} about the nature of the gap in \tise. By adsorbing water on the surface, Rasch \textit{et al.} \cite{Rasch2008} induced a band bending, the intensity of which is comparable to the energy position of the coherent peak above the Fermi level in the semiconductor case [Fig. \ref{fig:2}(d)]. Thus Rasch \textit{et al.} deduced a semiconducting behavior from the measurement of the coherent peak. On the other hand, Li \textit{et al.} \cite{Li2007a} measured a very low carrier density using optical spectroscopy and concluded that the compound is metallic. This very low carrier density at the Fermi level is related to the incoherent part crossing the Fermi level in the semiconductor case [Fig. \ref{fig:2}(d)]. In this sense, the final electronic structure, taking electron-hole effects into account, has, at the same time, a semiconductor and a semimetal character. 

This project was supported by the Fonds National Suisse (FNS) pour la Recherche Scientifique through Div. II. Skillful assistance was provided by Dr. Pawel Bednarek for the computer cluster calculations. C.M. acknowledges also support by the FNS under Grant No. PZ00P2\textunderscore154867.
\bibliographystyle{apsrev4-1} 
\bibliography{library}

\begin{thebibliography}{23}
\bibitem [{DFT()}]{DFTexpl}%
  \BibitemOpen
  \href@noop {} {\bibinfo  {journal} {We used the modified Becke-Johnson
  exchange potential with spin-orbit coupling. The fit is done near the high
  symmetry points over a $\Delta k = 0.1$ \AA $^{-1}$. The valence band is
  isotropic with an effective mass of $m_\textrm{VB} = 0.22$, the conduction
  band is forming an ellipse with $m_{\textrm{CB}x} = 4.3$ and
  $m_{\textrm{CB}y} = 0.29$. $x$ is the direction from the center of the BZ to
  its border and $y$ is the perpendicual in-plane direction. The masses are
  expressed in units of $m_0$. The relative errors are 10\%}\ }\BibitemShut
  {NoStop}%
\end{thebibliography}
\end{document}